\documentclass[conference]{IEEEtran}
\usepackage{amsmath}
\usepackage{times, epsfig}
\usepackage{color}
\usepackage{booktabs}
\usepackage{tabularx}
\usepackage{amsmath}

\UseRawInputEncoding
\begin{document}
\title{
System-level Simulation of Reconfigurable Intelligent Surface assisted Wireless Communications System
}

\author{
Qi Gu$^\dag$, Dan Wu$^\dag$, Xin Su$^\dag$, Hanning Wang$^\dag$, Jingyuan Cui$^\ast$, and Yifei Yuan$^\dag$\\
$^\dag$ Future Mobile Technology Laboratory, China Mobile Research Institute, Beijing, China\\
$^\ast$ School of Information and Communication Engineering, Beijing University of Posts and Telecommunications \\ 
 E-mail: $^\dag$\{guqi, wudan, suxin, wanghanning, yuanyifei\}@chinamobile.com,
$^\ast$ cjy123@bupt.edu.cn
}
\maketitle
\thispagestyle{empty}
\pagestyle{empty}

\begin{abstract}
Reconfigurable intelligent surface (RIS) is an emerging technique employing metasurface to reflect the signal from the source node to the destination node.
By smartly reconfiguring the electromagnetic (EM) properties of the metasurface and adjusting the EM parameters of the reflected radio waves, RIS can turn the uncontrollable propagation environment into an artificially reconfigurable space, and thus, can significantly  increase the communications capacity and improve the coverage of the system.
In this paper, we investigate the far field channel in which the line-of-sight (LOS) propagation is dominant.
We propose an antenna model that can characterize the radiation patterns of realistic RIS elements, and consider the
signal power received from the two-hop path through RIS.
System-level simulations of network performance under various scenarios and parameter configurations are performed, which show significant gain of RIS.
\end{abstract}


\section{Introduction} \label{sec:intro}
Metamaterials are a class of materials with artificially engineered structures and have extraordinary physical properties not found in natural materials.
The three-dimensional structure of metamaterials early age was hard to fabricate due to the challenges of complex structure, high dielectric loss, and difficult manufacturing, etc.
On the other hand, ultra-thin two-dimensional form of metamaterials, namely metasurface, becomes an alternative.
A metasurface is an artificial reflective surface made of elements orderly arranged on a plane. Incident ectromagnetic (EM) waves can be tuned by changing the metasurface's EM structure or arrangement of its elements.
Moreover, by introducing tunable components for each element, the metasurface structure becomes a more flexible, i.e.,  Reconfigurable Intelligent Surface (RIS).
It is consisted of a large-scale device array and an array control module \cite{wu2019towards}.
By properly adjusting each element on these large-scale device array, the response signals through each element can be constructively  superimposed to form a specific beam at the macroscopic scale.
The phase adjustment can be managed by the control module, which serves as the ``brain'' of the RIS.
In this way, the EM properties of the RIS and also the properties of reflected electromagnetic waves, including frequency-dependent phase and amplitude, can be manipulated to meet the requirement of wireless communications in real applications.

RIS consumes less energy and can easily be implemented under various scenarios. These features make it a strong candidate to enhance the end-to-end performance in a wireless communications system as a relay. According to  \cite{pan2021reconfigurable}, RIS is regarded as one of the most promising and revolutionary techniques for enhancing the spectral and/ or energy efficiency of wireless systems.
Moreover,
it is shown that RIS with large metasurface and massive elements can outperform active relay (whose power amplifier has higher energy consumption for forwarding signal from the source) in terms of spectral efficiency and energy efficiency \cite{bjornson2019intelligent,ntontin2020rate}.
RIS has triggered a research boom in the academia and industry, and attracted a lot of research attention in the society of communications and signal processing.
In \cite{gu2021performance}, a comprehensive comparison is made between RIS and half-duplex relay.
Besides, RIS has the advantage of capacity enhancement and coverage extension\cite{zhang2020capacity}.
In the case of coverage holes, RIS can be placed between BS and UE to provide supplementary links and improve the quality of communications \cite{han2019large}.

To study the RIS performance, it is necessary to model the channel model between BS and UE connected by RIS.
Rician fading channel model and a clear line-of-sight (LOS) path between BS-RIS as well as RIS-UE were considered in \cite{han2021dual}.
Channel models and the path loss exponent with single RIS and multiple RISs under Rician fading  were analyzed for indoor and outdoor non line-of-sight (NLOS) scenarios \cite{yildirim2020modeling}.
A clustered statistical MIMO model of RIS for millimeter wave (mmW) frequencies in indoor and outdoor environments were proposed by considering the multipath impact in the practical environment \cite{basar2021indoor}.
A geometric RIS-assisted multiple-input multiple-output (MIMO) channel model was proposed for fixed-to-mobile (F2M)  communications based on a three-dimensional cylinder model \cite{sun20213d}.
Currently, most of the studies have analytically demonstrated the advantages of RIS technology with single RIS or multiple RISs in single cell.
There is lack of system-level simulations on the multi-cell network when RIS is applied.
It is necessary to carry out system-level simulations over multiple cells to evaluate the overall performance with multiple RISs in real scenarios.

In this paper, the channel model for far field LOS dominant condition are investigated for the performance evaluation of the wireless networks with RIS.
A new antenna model is proposed to capture the characteristic of realistic RIS.
We focus on pathloss and large-scale fading and consider the long-term signal power received from the two-hop transmission relayed by RIS.
System-level simulations of network performance under various scenarios and parameter configurations are conducted based on the antenna model of RIS.
Conclusions and strategies related to RIS deployment are provided in the end.

\section{System Model}
In this section, system model will be introduced, including antenna model of RIS elements, channel model of a RIS-assisted system and the optimal phase of RIS.
Notations and abbreviations used in this work are listed in Table \ref{ta:notation}.

\begin{table}[!htb]
  \centering
    \caption{Notation Used}
    \begin{tabular}{lp{6cm}}
    \toprule
     \textbf{Symbols} & \textbf{Description} \\
     \midrule
      $F_{\theta}(\theta,\phi)$ & the antenna element field pattern (amplitude) in the vertical polarization\\
      \hline
      $F_{\phi}(\theta,\phi)$ & the antenna element field pattern (amplitude) in the horizontal polarization\\
      \hline
      $\bar{d}_{l,k}$ & local coordinate of the $k$-th element of RIS $l$ \\
      \hline
      $K$ & the number of RIS elements\\
      \hline
      $P_{RIS_l}$ & the received power received via RIS $l$\\
      \hline
      $PL_{BS-RIS_l}$ & pathloss from BS to RIS $l$  \\
      \hline
      $PL_{RIS_l-UE}$ & pathloss from RIS $l$ to UE\\
      \hline
      $\hat{r}_{ZOA, AOA}$ & spherical unit vector with azimuth arrival angle and elevation arrival angel\\
      \hline
      $\hat{r}_{ZOD, AOD}$ & spherical unit vector with azimuth departure angle and elevation departure angel\\
      \hline
      $S$ & the number of transmit antenna elements of BS \\
      \hline
      $SF_{BS-RIS_l}$ & shadow fading from BS to RIS $l$ \\
      \hline
      $SF_{RIS_l-UE}$ & shadow fading from RIS $l$ to UE \\
      \hline
      $TX_{power}$ & the transmitted power of BS \\
      \hline
      $U$ & the number of receive antenna elements of UE\\
      \hline
      $w_s$ & a complex weight vector used for virtualization port\\
      \hline
      $\Phi_{l,k}$ &  reflection coefficients of the $k$-th element of RIS $l$\\
      \hline
      $\alpha_{1,l,k}^{far} $ & phase factor of channel coefficients from BS to the $k$-th element of RIS $l$\\
      \hline
      $\alpha_{2,l,k,u}^{far} $ & phase factor of channel coefficients from the $k$-th element of RIS $l$to the u-th receive antenna of UE\\
      \hline
      $\theta$ & zenith angle\\
      \hline
      $\phi$&   azimuth angle \\
      \hline
      $\lambda$ &  wavelength\\
      \bottomrule
     \toprule
      \textbf{Abbreviations} &  \\
     \midrule
    AOA& Azimuth angle Of Arrival\\
     \hline
    AOD& Azimuth angle Of Departure\\
    \hline
     BS& Base Station\\
     \hline
     UE& User Equipment\\
    \hline
    ZOA& Zenith angle Of Arrival\\
    \hline
    ZOD& Zenith angle Of Departure\\
    \bottomrule
    \end{tabular}\label{ta:notation}
\end{table}

\subsection{Antenna Model of RIS Elements}\label{sec:antenna_model}
The antenna pattern of active radiating element in the existing model TR 38.901 Table 7.3-1 \cite{3GPP2019} is shown as Table \ref{ta:single_antenna_38901}.
\begin{table*}[!htb]
  \centering
    \caption{Radiation power pattern of a single antenna element}
    \begin{tabular}{lp{8.8cm}}
    \toprule
     \textbf{Parameter} & \textbf{Values} \\
     \midrule
      Vertical cut of the radiation power pattern (dB)  & $A_{\text{dB}}(\theta,\phi=0^{\circ})=-\min \left\{ 12\left( \frac{\theta-90^{\circ}}{\theta_{3\text{dB}}}\right)^2,SLA_V \right\}$
      \\& with $\theta_{3\text{dB}}=65^{\circ}, SLA_V=30\text{dB}$  and $\theta \in [0^{\circ},180^{\circ}]$  \\
      \hline
      Horizontal cut of the radiation power pattern (dB) & $A_{\text{dB}}(\theta=90^{\circ},\phi)=-\min \left\{ 12\left( \frac{\phi}{\phi_{3\text{dB}}}\right)^2, A_{\text{max}} \right\}$
      \\&  with $\phi_{\text{3dB}}=65^{\circ}, A_{\text{max}}=30\text{dB}$  and $\phi \in [-180^{\circ},180^{\circ}]$\\
      \hline
      3D radiation power pattern (dB)& $A_{\text{dB}}(\theta,\phi)=-\min \{ -(  A_{\text{dB}}(\theta,\phi=0^{\circ}) +  A_{\text{dB}}(\theta=90^{\circ},\phi) ), A_{\text{max}}\}$ \\
      \hline
      Maximum directional gain of an antenna element $G_{\text{E,max}}$ & $8$ dBi\\
      \bottomrule
    \end{tabular}\label{ta:single_antenna_38901}
\end{table*}
Given the zenith angle $\theta$ and the azimuth angle $\phi$, the power pattern of antenna element can be defined.
However, since RIS is a passive metasurface, its antenna model is different from the active radiating element£ºthe incident angle is not always along the normal direction of RIS panel. Also RIS element can be modeled as a single radiating element without directional gain, i.e.
 $G_{\text{E,max}}=0$ dBi.

\subsection{Channel Model}
To our best knowledge, there is no comprehensive modeling of RIS channels so far.
Hence, the extension over the existing channel model seems to be a pragmatic approach in this case.
The most widely used model is the channel model defined in 3GPP TR 38.901.
As the RIS can be considered as a node in the far-field assumption and is expected to be deployed higher than the UE in outdoor scenarios, many aspects such as pathloss, antennas polarization and the phase difference at different positions of a continuous surface and so on, are similar to these for the channel model in 3GPP TR 38.901.
Therefore, our model is based on 3GPP TR 38.901 which is applied in both BS-RIS and RIS-UE links.
To simplify the model, we consider the far-field conditions and LOS path.

The received power at UE via the RIS-assisted two-hop path can be represented as Eq. \eqref{eq:RIS_channel_model} that incorporated path loss ($PL$), shadow fading ($SF$), phase factor ($\alpha$) of two-hop and phase compensation of RIS ($\Phi_{l,k}$), where
$F_{\theta}(\theta_{ZOA_{RIS}},\varphi_{AOA_{RIS}})$, $F_{\varphi}(\theta_{ZOA_{RIS}},\varphi_{AOA_{RIS}})$, $F_{\theta}(\theta_{ZOD_{RIS}},\varphi_{AOD_{RIS}})$ and $F_{\varphi}(\theta_{ZOD_{RIS}},\varphi_{AOD_{RIS}})$ in Eq. \eqref{eq:RIS_channel_model_b} and Eq. \eqref{eq:RIS_channel_model_c} represent the power pattern of a RIS element.

\begin{figure*}[!htb]
\begin{subequations}
\begin{align}
&P_{RIS_l} = PL_{BS-RIS_{l}}\cdot PL_{RIS_{l}-UE}\cdot SF_{BS-RIS_{l}} \cdot SF_{RIS_{l}-UE}
  \sum_{u=1}^{U} \left|  \sum_{k=1}^{K}   \alpha_{2,l,k}^{far} \cdot
e^{j\Phi_{l,k}} \cdot \alpha_{1,l,k}^{far}   \right|^2   \cdot \frac{TX_{power}}{U},  \label{eq:RIS_channel_model_a}\\
\text{where}\nonumber\\
&
\alpha_{1,l,k}^{far} =
\underbrace{\left [
\begin {array} {c}
F_{\theta}(\theta_{ZOA_{RIS}},\varphi_{AOA_{RIS}})\\
F_{\varphi}(\theta_{ZOA_{RIS}},\varphi_{AOA_{RIS}})
\end {array}
\right ]^T
}_{\text{the pattern of RIS}}
\left [
\begin {array} {cc}
1 & 0\\
0 & -1
\end {array}
\right ]
\underbrace{
\left [
\begin {array} {c}
F_{\theta}(\theta_{ZOD_{BS}},\varphi_{AOD_{BS}})\\
F_{\varphi}(\theta_{ZOD_{BS}}\varphi_{AOD_{BS}})
\end {array}
\right ]
}_{\text{the pattern of antenna element of BS}}
\cdot
\text{exp}\left(j 2 \pi  \frac{\hat{r}^T_{ZOA_{RIS},AOA_{RIS}}\cdot\bar{d}_{l,k}}{\lambda} \right)
,\label{eq:RIS_channel_model_b}\\
&
\alpha_{2,l,k,u}^{far} =
\underbrace{
\left [
\begin {array} {c}
F_{\theta}(\theta_{ZOA_{UE}},\varphi_{AOA_{UE}})\\
F_{\varphi}(\theta_{ZOA_{UE}},\varphi_{AOA_{UE}})
\end {array}
\right ]^T
}_{\text{the pattern of antenna element of UE}}
\left [
\begin {array} {cc}
1 & 0\\
0 & -1
\end {array}
\right ]
\underbrace{
\left [
\begin {array} {c}
F_{\theta}(\theta_{ZOD_{RIS}},\varphi_{AOD_{RIS}})\\
F_{\varphi}(\theta_{ZOD_{RIS}},\varphi_{AOD_{RIS}})
\end {array}
\right ]
}_{\text{the pattern of RIS}}
\cdot
\text{exp}\left(j 2 \pi  \frac{\hat{r}^T_{ZOD_{RIS},AOD_{RIS}}\cdot\bar{d}_{l,k}}{\lambda} \right), \label{eq:RIS_channel_model_c}
\end{align}
with
\begin{align}
&\hat{r}^T_{ZOA_{RIS},AOA_{RIS}}=\left [
\begin {array} {c}
\sin \theta_{ZOA_{RIS}}\cos \varphi_{AOA_{RIS}}\\
\sin \theta_{ZOA_{RIS}}\sin \varphi_{AOA_{RIS}}\\
\cos \theta_{ZOA_{RIS}}
\end {array}
\right ],
\quad
\hat{r}^T_{ZOD_{RIS},AOD_{RIS}}=\left [
\begin {array} {c}
\sin \theta_{ZOD_{RIS}}\cos \varphi_{AOD_{RIS}}\\
\sin \theta_{ZOD_{RIS}}\sin \varphi_{AOD_{RIS}}\\
\cos \theta_{ZOD_{RIS}}
\end {array}
\right ],
\end{align}
\begin{align} \label{eq:F_BS1}
F_{\theta}(\theta_{ZOD_{BS}},\varphi_{AOD_{BS}})
= \sum_{s=1}^{S} w_s \
\text{exp}\left(j 2 \pi  \frac{\hat{r}^T_{\theta_{ZOD_{BS,s}},\varphi_{AOD_{BS,s}}} \cdot\bar{d}_s}{\lambda_0} \right) \cdot F_{\theta}(\theta_{ZOD_{BS,s}},\varphi_{AOD_{BS,s}}),
\end{align}
\begin{align} \label{eq:F_BS2}
F_{\varphi}(\theta_{ZOD_{BS}},\varphi_{AOD_{BS}})
= \sum_{s=1}^{S} w_s \
\text{exp}\left(j 2 \pi  \frac{\hat{r}^T_{\theta_{ZOD_{BS,s}},\varphi_{AOD_{BS,s}}} \cdot\bar{d}_s}{\lambda_0} \right) \cdot F_{\varphi}(\theta_{ZOD_{BS,s}},\varphi_{AOD_{BS,s}}),
\end{align}
\begin{align}
\hat{r}_{\theta_{ZOD_{BS,s}},\varphi_{AOD_{BS,s}}} =
\left [
\begin {array} {c}
\sin \theta_{ZOD_{BS,s}}  \cos \varphi_{AOD_{BS,s}} \\
\sin \theta_{ZOD_{BS,s}}  \sin \varphi_{AOD_{BS,s}} \\
\cos \theta_{ZOD_{BS,s}}
\end {array}
\right ].
\end{align}\label{eq:RIS_channel_model}
\end{subequations}
\hrulefill
\end{figure*}

\subsection{The Optimal Phase of the RIS Element}
Let us define the exponent term in Eq. \eqref{eq:RIS_channel_model_b} and Eq. \eqref{eq:RIS_channel_model_c} as
\begin{align}
\tilde{\Phi}_{l,k}= & 2 \pi \frac{(\hat{r}^T_{ZOA_{RIS},AOA_{RIS}}+\hat{r}^T_{ZOD_{RIS},AOD_{RIS}})\cdot\bar{d}_{l,k}}{\lambda},
\end{align}
which is the phase accrued by propagation over the path from BS-RIS and RIS-UE including the $k$-th RIS element.
In Eq. \eqref{eq:RIS_channel_model_a}, the optimal phase $\Phi_{l,k}$ of $k$-th element of RIS $l$ is complex-valued coefficient of $\tilde{\Phi}_{l,k}$, i.e.
\begin{align}
\Phi_{l,k}= - \tilde{\Phi}_{l,k},
\label{eq:optimal_phase}
\end{align}
which is the phase difference at different positions of a continuous surface.
It is worthy to note that the optimal phase $\Phi_{l,k}$ is not quantized.
Apparently RIS here is used to compensate for phase difference.

\subsection{Verification}

\begin{figure}[htbp]
\centering
\begin{minipage}[t]{0.25\textwidth}
\centering
\includegraphics[height=33mm,width=46mm]{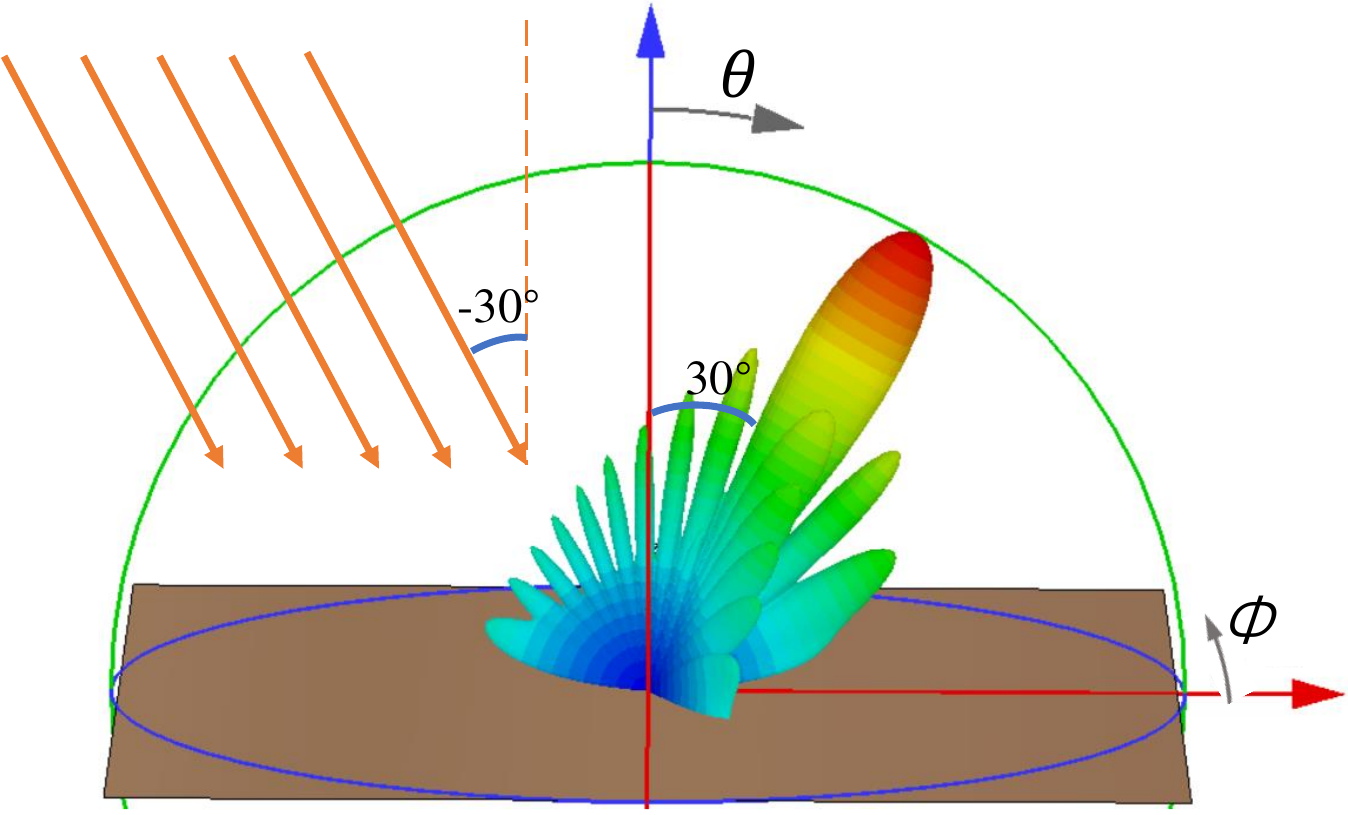}
\label{fig:new_pattern}
\end{minipage}
 \begin{minipage}[t]{0.2\textwidth}
\centering
\includegraphics[height=42mm,width=40mm]{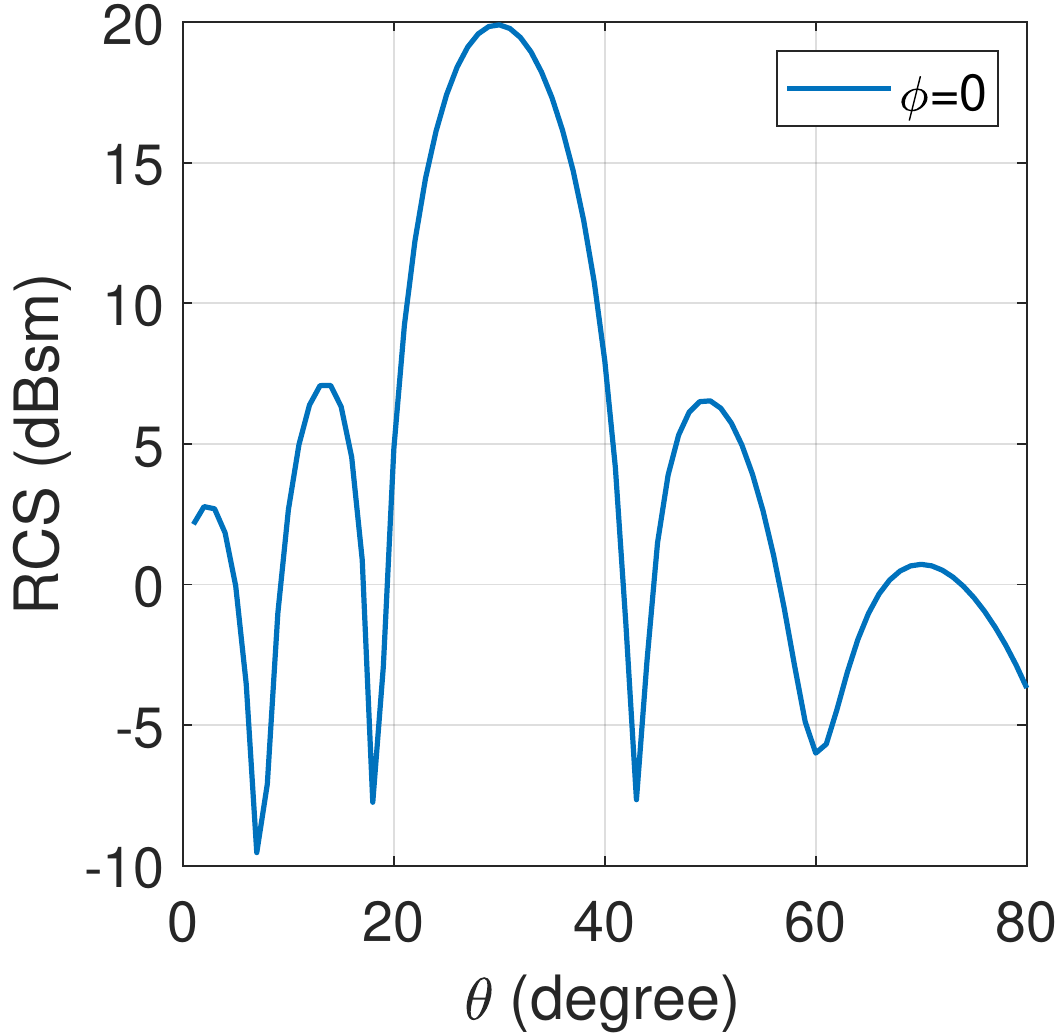}
\label{fig:new_pattern2}
\end{minipage}
\caption{Radiation power pattern of a metasurface obtained by RIS hardware EM wave simulation}
\label{fig:fig_power_pattern_new}
\end{figure}

The passive characteristics of the RIS has been verified by the RIS hardware simulations using the commercial electromagnetic solver (e.g. CST Studio Suite), as shown in Fig. \ref{fig:fig_power_pattern_new}.
For an input wave with an incident angle of 30 degree, when the phase compensations of RIS elements are all zeros (e.g. pure reflector) the wave is reflected by a metasurface, where the main lobe direction of the outgoing wave is 30 degree.
By observing on the direction of outgoing wave's main lobe by changing the wave's incident angle,
it is found that the power pattern of passive RIS element is related to the angle of the incident wave, the direction of outgoing wave's main lobe is symmetrical with the incident wave about the normal direction.

\begin{figure}[htbp]
\centering
\begin{minipage}[t]{0.25\textwidth}
\centering
\includegraphics[height=38mm,width=30mm]{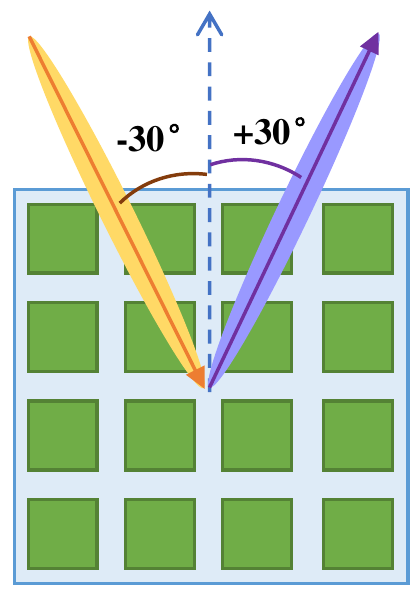}
\label{fig_new_pattern_3030}
\end{minipage}
 \begin{minipage}[t]{0.2\textwidth}
\centering
\includegraphics[height=43mm,width=40mm]{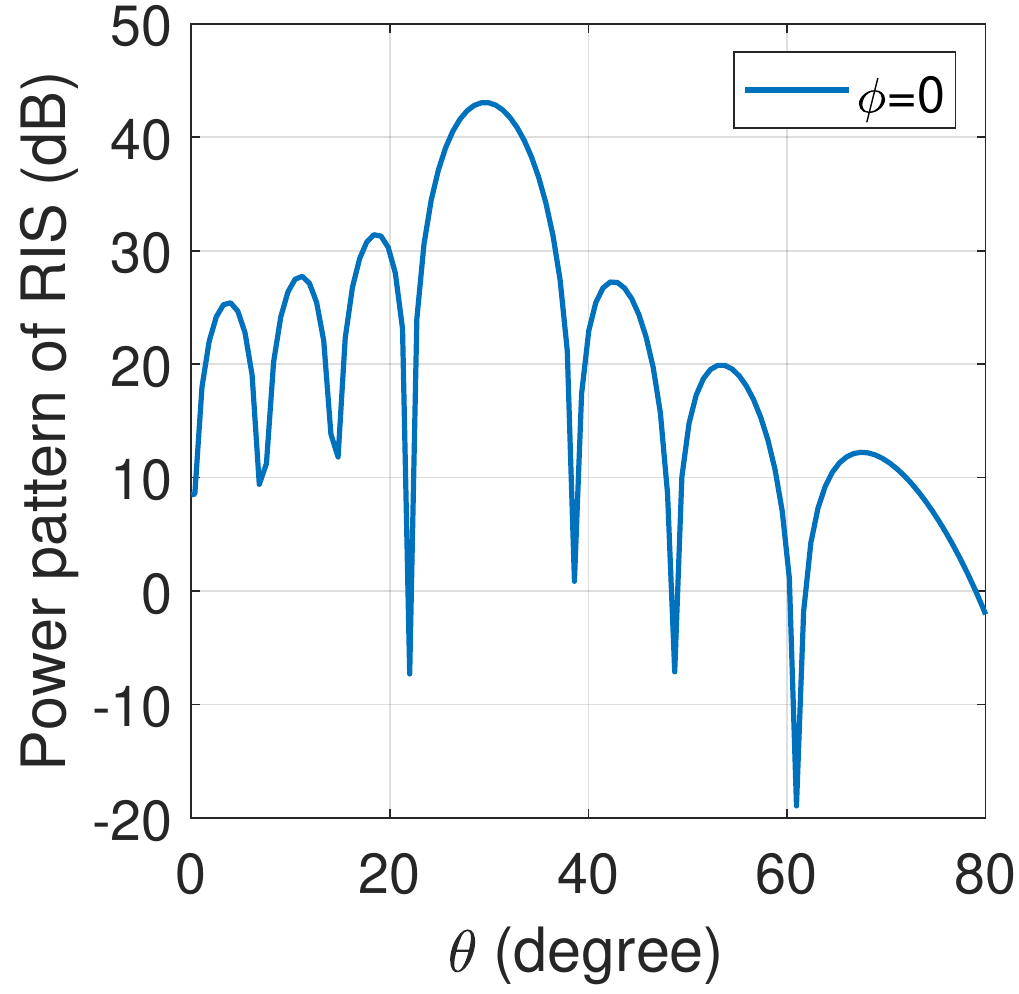}
\label{fig_new_pattern_3030_matlab}
\end{minipage}
\caption{Power pattern of passive RIS (without phase adjustment) by extending TR 38.901 model}
\label{fig:fig_power_pattern_3030}
\end{figure}

\begin{figure}[htbp]
\centering
\begin{minipage}[t]{0.25\textwidth}
\centering
\includegraphics[height=35mm,width=33mm]{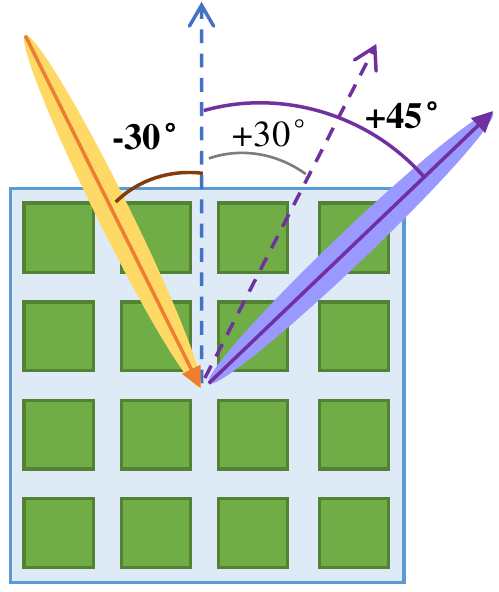}
\label{fig_new_pattern_3045}
\end{minipage}
 \begin{minipage}[t]{0.2\textwidth}
\centering
\includegraphics[height=43mm,width=40mm]{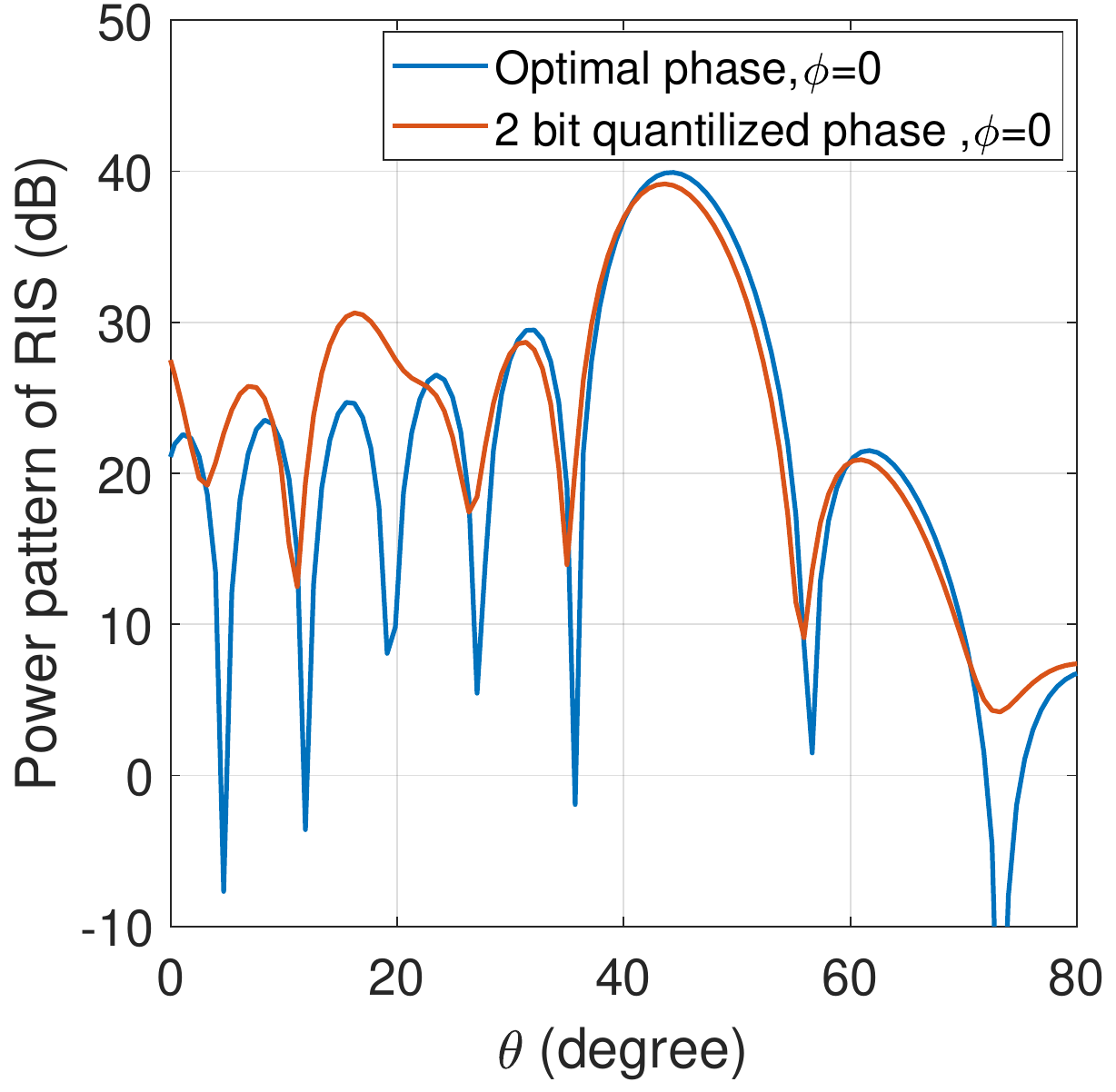}
\label{fig_new_pattern_3045_matlab}
\end{minipage}
\caption{Power pattern of passive RIS with phase adjustment}
\label{fig:fig_power_pattern_3045}
\end{figure}

By utilizing the power pattern of RIS proposed in Eq. \eqref{eq:RIS_channel_model_b} and Eq. \eqref{eq:RIS_channel_model_c}, as shown in Fig. \ref{fig:fig_power_pattern_3030}, for the incident wave with an incident angle of 30 degree, the main lobe direction of the outgoing wave is 30 degree without phase adjustment.
As shown in Fig. \ref{fig:fig_power_pattern_3045}, for the incident wave with an incident angle of 30 degree, a reflection angle of the outgoing wave can be 45 degree by phase adjustment in Eq. \eqref{eq:optimal_phase}.
It demonstrates the validity of the proposed antenna model of RIS.

\subsection{Impact of Bit Quantization}
In the theoretical analysis, the continuous phase adjustment of the RIS element is often assumed. However, due to the hardware limitations, continuous phase is infeasible in real implementation.
The design of the hardware and control modules of the RIS can be greatly simplified by discrete phases.
Phase quantization would cause certain performance loss compared to the continuous phase control.
In \cite{wu2019beamforming}, it shows that 1-bit, 2-bit and 3-bit discrete phases would result in power losses of 3.9 dB, 0.9 dB, and 0.2 dB, respectively, compared with continuous phases.
In \cite{han2019large}, compared with the optimal phase, the ergodic spectral efficiency of the 2 bit quantized phase decreases by less than 1 bit/s/Hz, the ergodic spectral efficiency of the 3 bit quantized phase is close to the performance of the optimal phase.
For a 2 bit discrete phase control of  RIS, the ideal compensation phase is mapped to discrete phases, for example:
\begin{align}
\Phi_{l,k}=
\begin{cases}
\pi/4, \quad &0 \leq \Phi_{l,k}\mod 2\pi  < \pi/2\\
3\pi/4, \quad  &\pi/2 \leq \Phi_{l,k}\mod 2\pi  < \pi\\
-3\pi/4, \quad &\pi \leq \Phi_{l,k}\mod 2\pi  < 3\pi/2\\
-\pi/4, \quad  & 3\pi/2 \leq  \Phi_{l,k}\mod 2\pi <2\pi
\end{cases}\label{eq:2bit_quantized_phase}
\end{align}
Fig. \ref{fig:fig_power_pattern_3045} shows the power pattern with optimal phase proposed in Eq. \eqref{eq:optimal_phase}  and 2 bit quantized phase in Eq. \eqref{eq:2bit_quantized_phase}, respectively.
It can be seen that the power pattern with 2 bit quantized phase is close to the power pattern with optimal phase.
In the practical application of RIS, the use of 2 bit quantization phase can meet the basic requirements.

\section{Simulation Results and Analysis}
In this section, we first describe the  system setup for system-level simulation, followed by simulation results and analysis.

\subsection{System Setup}
The system-level simulation platform used in this paper is implemented in C++ and contains more than several hundred classes and tens of thousand lines of code in total, which has been developed for years.
The platform is developed base on 5G system simulation that contains: 1) network topology ; 2) antenna pattern; 3) large/small scale channel model; 4) traffic types and service load models; 5) channel measurement and feedback process; 6)  resource scheduling and uplink power control process; 7) SINR statistic, throughput, delay, spectral efficiency and other performance indicators.
The simulator can fully support the performance study in 3GPP of various physical layer technologies.

\begin{figure*}[htbp]
\centering
\begin{minipage}[t]{0.32\textwidth}
\includegraphics[height=55mm,width=60mm]{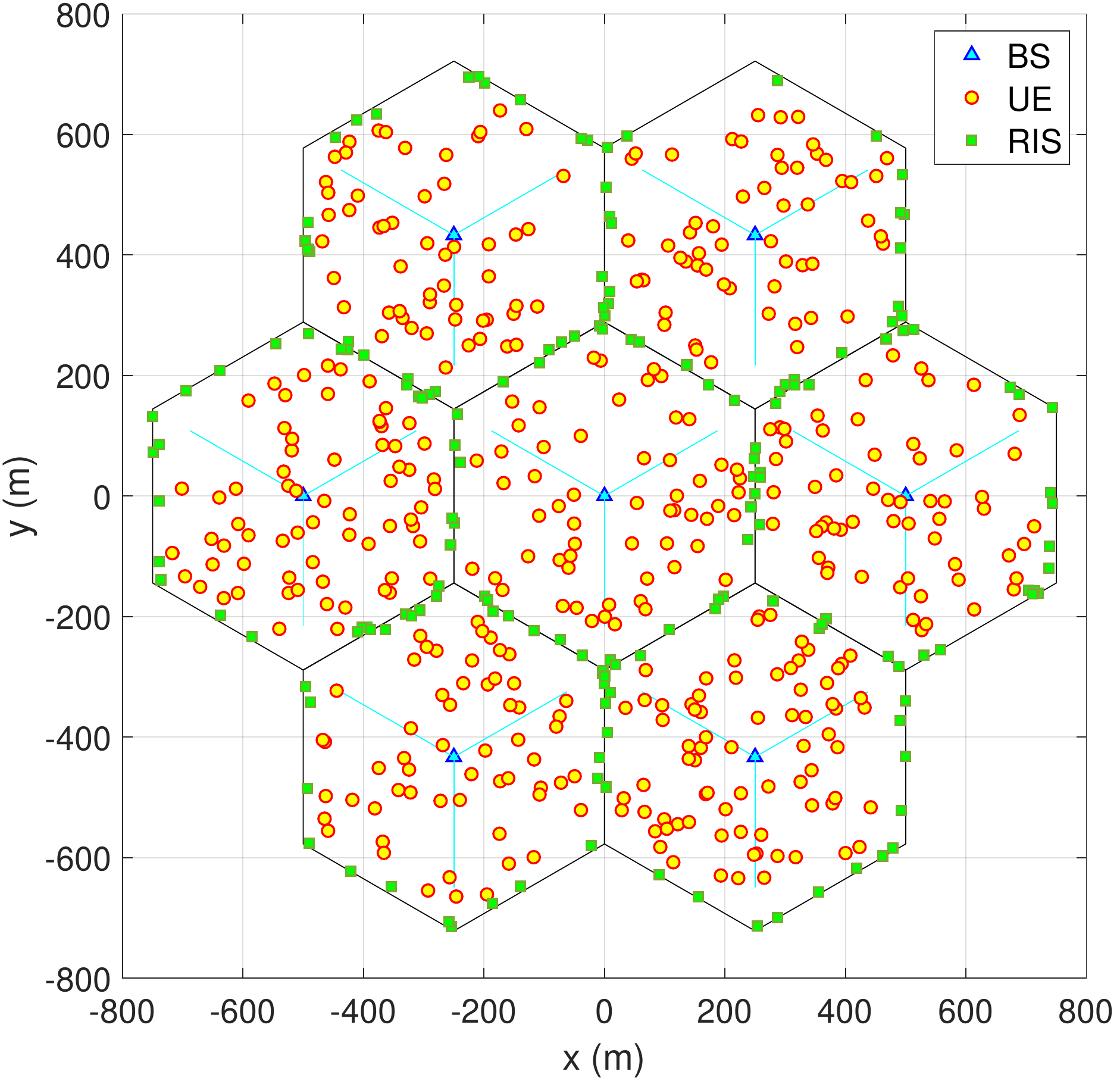}
\caption{UE randomly dropped, RIS deployed at cell edges}
\label{fig:fig_layout_UErandom_RISedge}
\end{minipage}
\begin{minipage}[t]{0.32\textwidth}
\centering
\includegraphics[height=55mm,width=60mm]{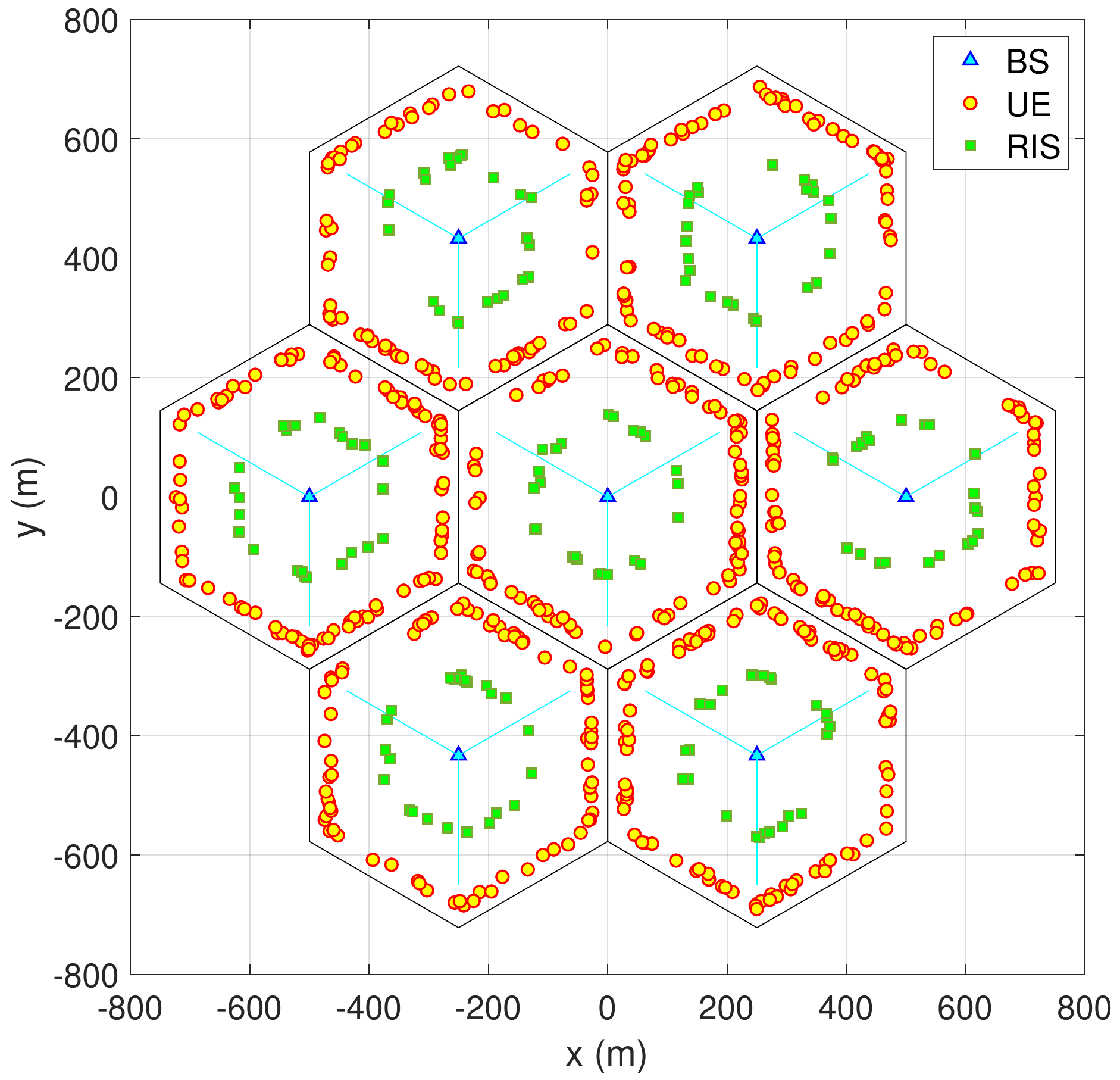}
\caption{RIS deployed in the middle of the cell, all UE at cell edges.}
\label{fig:fig_layout_UEedge_RISmiddle}
\end{minipage}
\begin{minipage}[t]{0.32\textwidth}
\centering
\includegraphics[height=55mm,width=60mm]{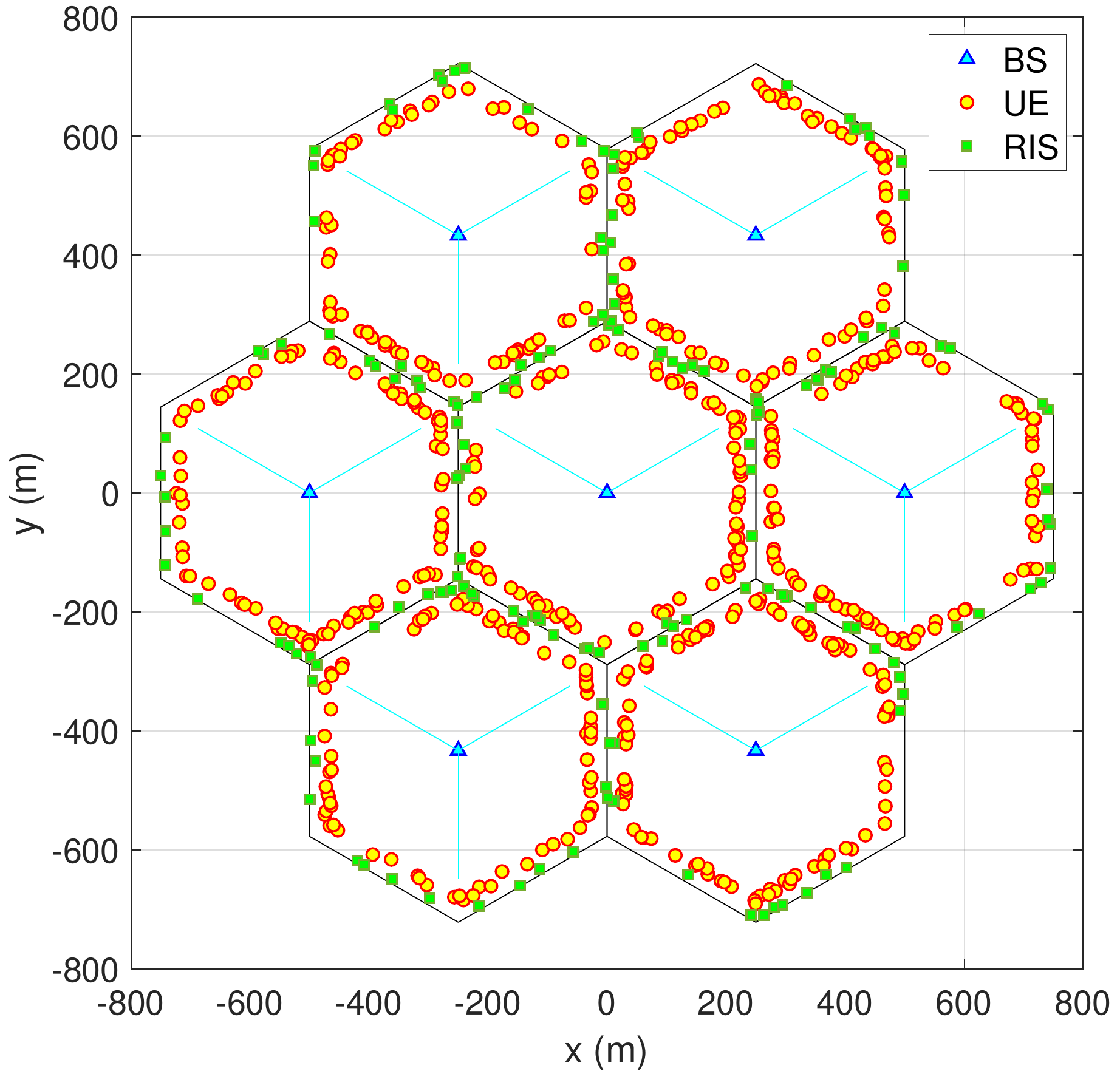}
\caption{RIS deployed at the edge of the cell, all UE at cell edges.}
\label{fig:fig_layout_UEedge_RISedge}
\end{minipage}
\end{figure*}

The system configuration and parameters are listed in Table \ref{ta:simulation}.
The scenario is shown as Fig. \ref{fig:fig_layout_UErandom_RISedge}, consisting of 7 cells and 21 sectors with a base station site-to-site of 500 m.
The antenna model of RIS is as described in Section \ref{sec:antenna_model}.
The pathloss model in 3GPP 38.901 Urban Macro (Uma) channel model \cite{3GPP2019} is used.
Here, we consider the large scale model.
Both the BS-RIS link and the RIS-UE link are LOS paths.
At 2.6 GHz, the heights of the base station, users and RIS are 25 m, 1.5 m and 15 m, respectively.
A horizontal spacing of half wavelength and a vertical spacing of 0.8 times the wavelength are assumed for the RIS elements.
The number of RIS  per cellular sector is $8$ or $16$ and the number of elements per RIS is $16*16=256$ or $40*40=1600$.


\begin{table}[!htb]
  \centering
    \caption{The Value of Parameters}
    \begin{tabular}{ll}
    \toprule
     \textbf{Parameters} & \textbf{Values} \\
     \midrule
      Downtilt of BS $h_{BS}$ & $0^{\circ}$ \\
      \hline
      Downtilt of RIS $h_{RIS}$ & $10^{\circ}$  \\
      \hline
      Frequency $f_c$ & 2.6 GHz \\

      \hline
      Height of BS $h_{BS}$ & 25 m \\
      \hline
      Height of RIS $h_{RIS}$ & 15 m \\
      \hline
      Height of UE $h_{UE}$ & 1.5 m \\
      \hline
      Horizonal spacing for the RIS elements & $0.5 \lambda$  \\
      \hline
      Number of cells and sectors & 7 cells and 21 sectors\\
      \hline
      Number of elements per RIS & $16*16$ or $40*40$ \\
      \hline
      Number of RISs per sector & 8 or 16 \\
      \hline
      Polarization & single-polarization\\
      \hline
      Vertiacal spacing for the RIS elements & $0.8 \lambda$  \\
      \bottomrule
    \end{tabular}\label{ta:simulation}
\end{table}

\subsection{Simulation Results}
In this subsection, simulation results are presented to demonstrate the performance of RIS.

\subsubsection{Users are randomly distributed at the cell}
Firstly, users are randomly dropped within 7 cells and RIS deploy at the edge of cell as shown in Fig.  \ref{fig:fig_layout_UErandom_RISedge}.

\begin{figure}[htbp]
\centering
\includegraphics[height=56mm,width=56mm]{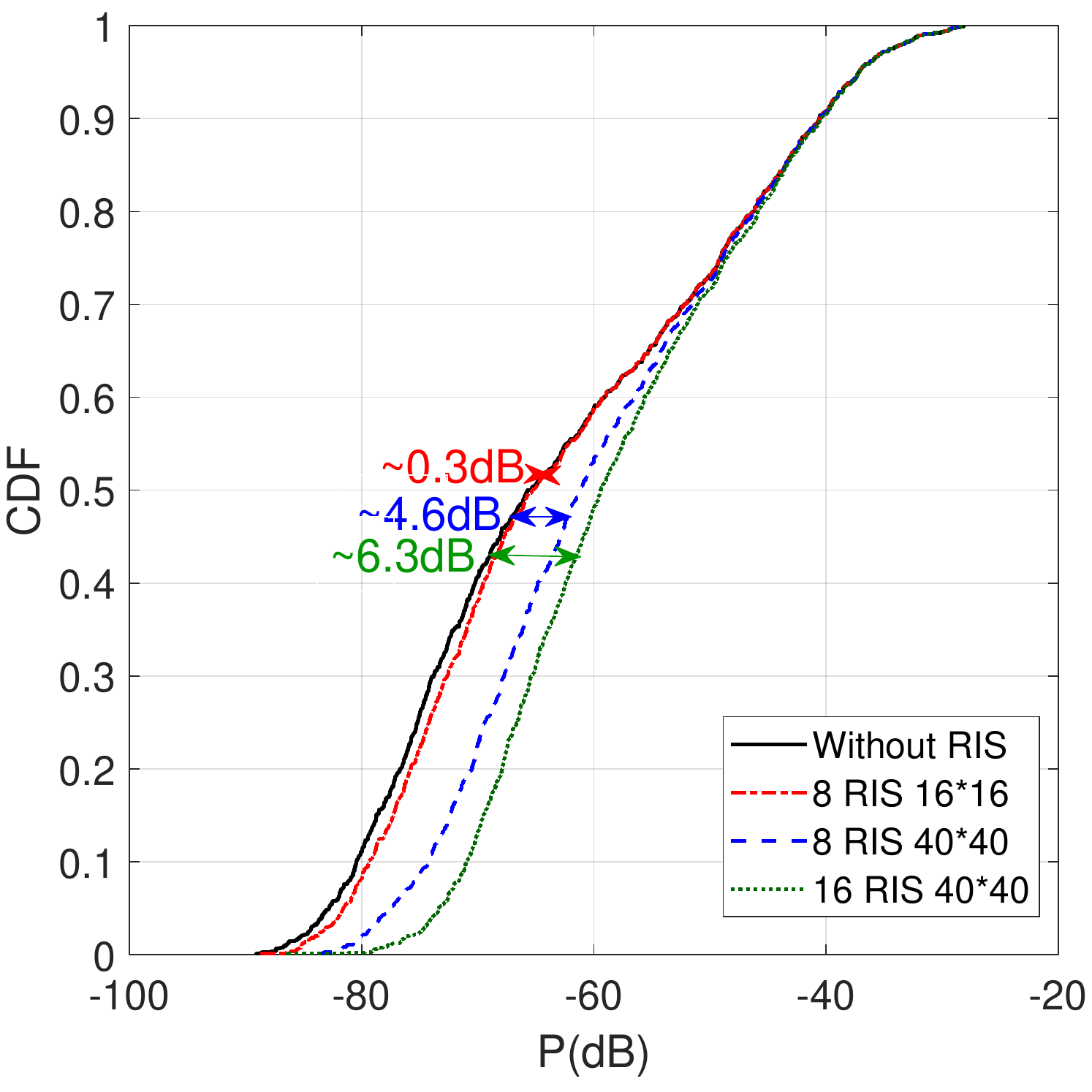}
\caption{CDF of the received signal power $P$ when UE randomly dropped, RIS deployed at edges}
\label{fig:fig_simulation_1_Pt}
\end{figure}

\begin{figure}[htbp]
\centering
\includegraphics[height=56mm,width=56mm]{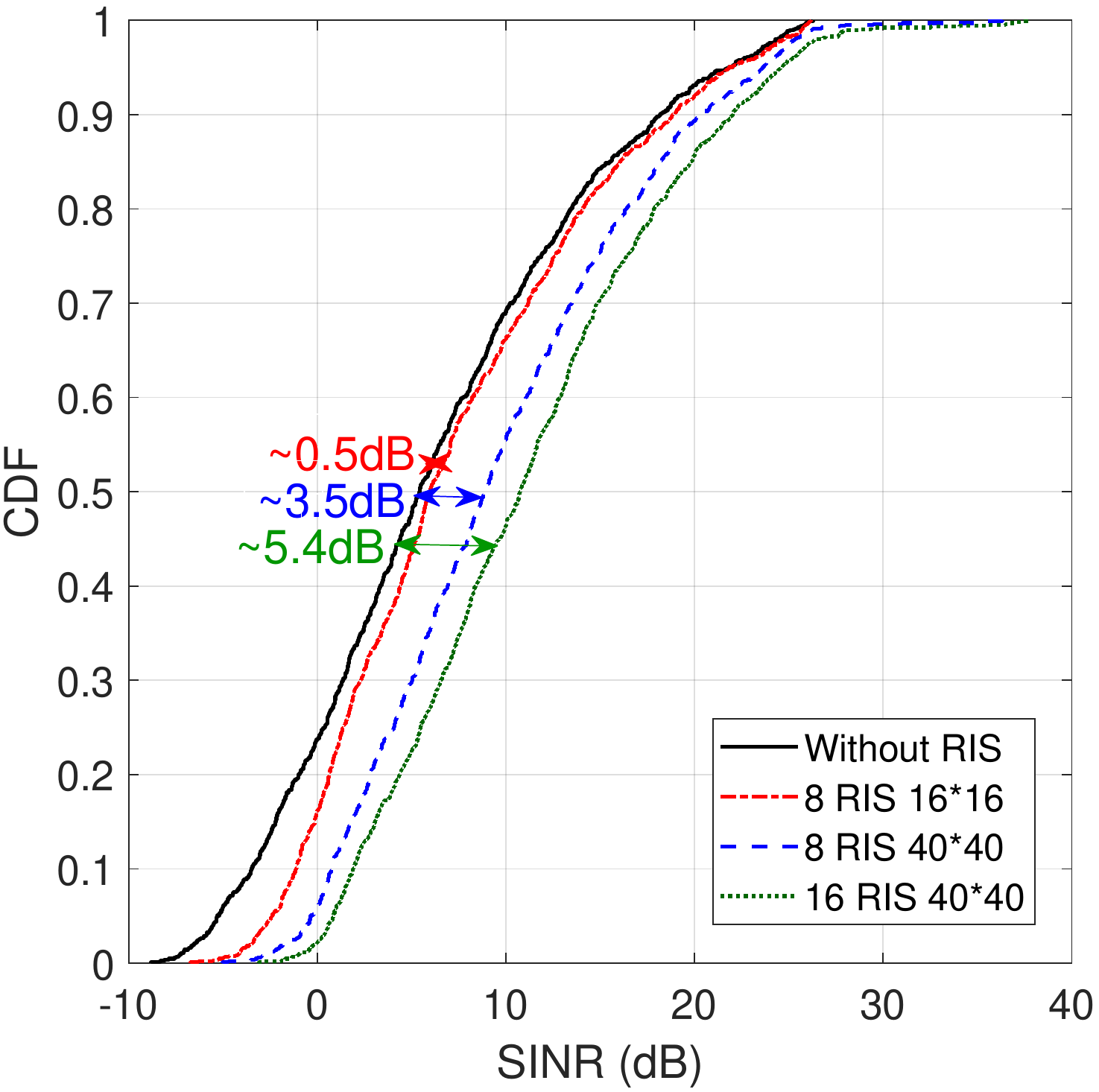}
\caption{CDF of SINR when UE randomly dropped, RIS deployed at edges}
\label{fig:fig_simulation_1_SINR}
\end{figure}

Fig. \ref{fig:fig_simulation_1_Pt} shows the cumulative distribution function (CDF) of the received signal power $P$.
Compared to the case without RIS, the received signal power gains are about 0.3 dB, 4.6 dB, and 6.3 dB for 8 RIS panels, each having  16*16 elements, 8 RIS panels, each having 40*40 elements, and 16 RIS panels, each having  with 40*40 elements, respectively.
Fig. \ref{fig:fig_simulation_1_SINR} shows the CDF of signal to interference plus noise ratio (SINR).
The interference counted in this paper includes three types: the direct path interference of the neighboring base station, the interference of the neighboring base station through the RIS panels  of the neighboring cell, and the interference of the neighboring base station through the RIS panels of this cell.
Compared to the case without RIS, the SINR gains are about 0.5 dB, 3.5 dB, and 5.4 dB for 8 RIS with 16*16 elements, 8 RIS with 40*40 elements, and 16 RIS with 40*40 elements, respectively.
The simulation results show the performance gain improved by deploying RIS in the cell. The system performance increases as either the number of elements per RIS or the number of RIS per sector increases.

\begin{figure}[htbp]
\centering
\includegraphics[height=56mm,width=56mm]{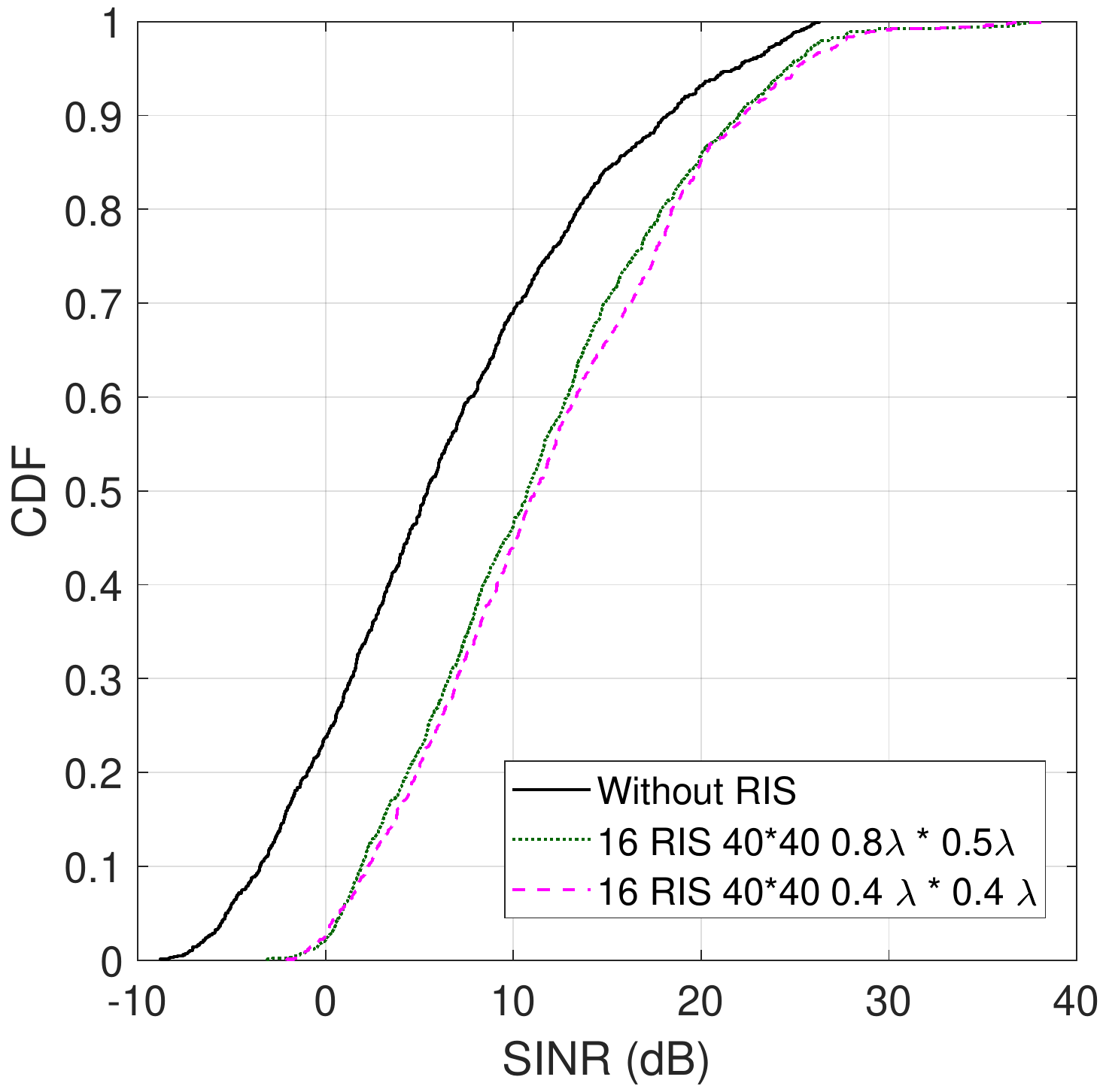}
\caption{CDF of SINR when UE randomly dropped, RIS with element spacing $ 0.8\lambda*0.5 \lambda$ and $ 0.4\lambda*0.4 \lambda$ deployed at edges}
\label{fig:fig_simulation_1_SINR_diffelementdistance}
\end{figure}

Fig. \ref{fig:fig_simulation_1_SINR_diffelementdistance} plots the CDF of SINR with element spacing $ 0.8\lambda*0.5 \lambda$ and $ 0.4\lambda*0.4 \lambda$.
Mutual coupling between the elements are not considered.
It shows that RIS with smaller element spacing can contribute to higher performance.

\subsubsection{Users are distributed at the cell edges}
\begin{figure}[htbp]
\centering
\includegraphics[height=56mm,width=56mm]{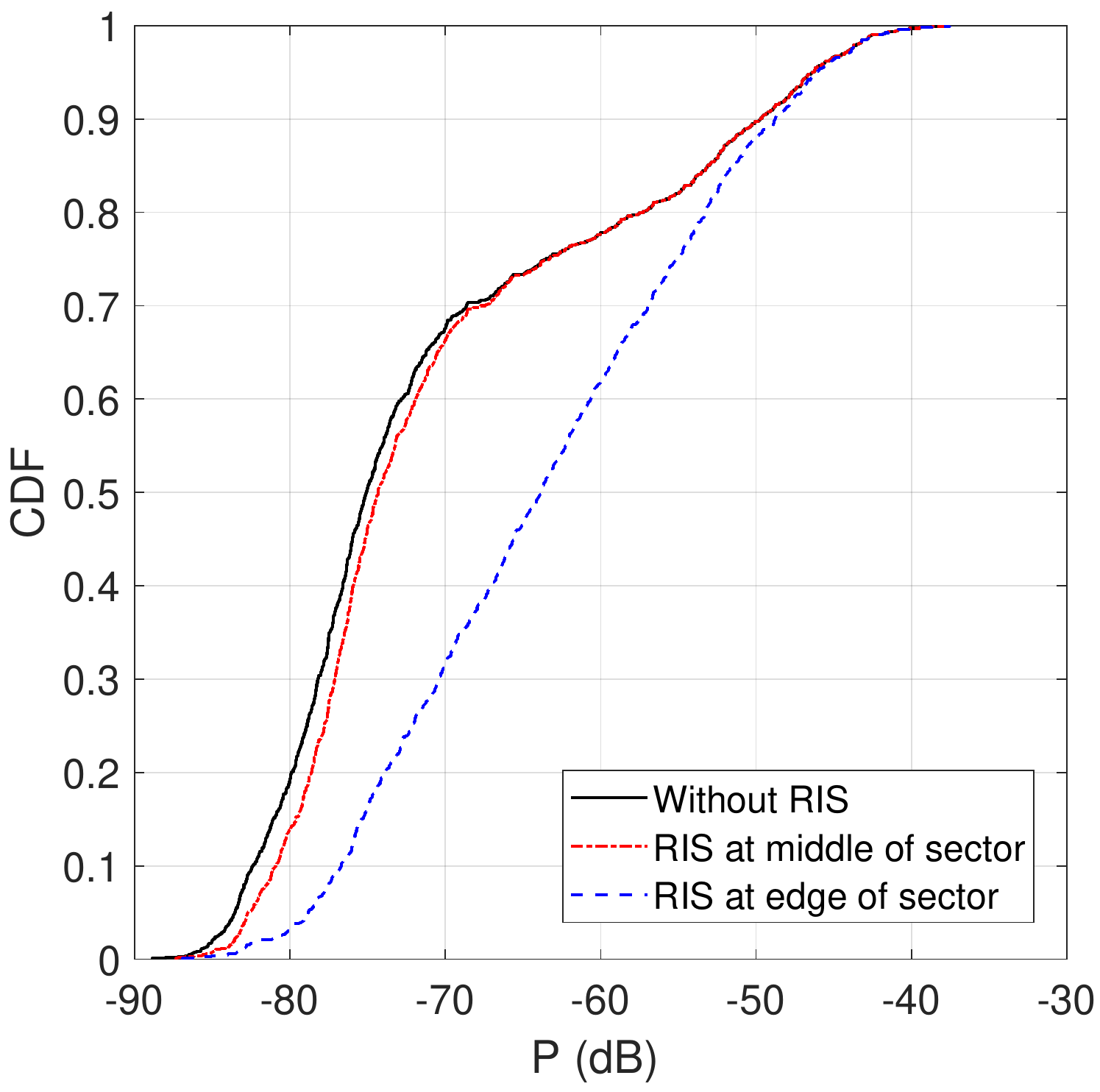}
\caption{CDF of the received signal power $P$, all UE at cell edges }
\label{fig:fig_simulation_2_Pt}
\end{figure}

\begin{figure}[htbp]
\centering
\includegraphics[height=56mm,width=56mm]{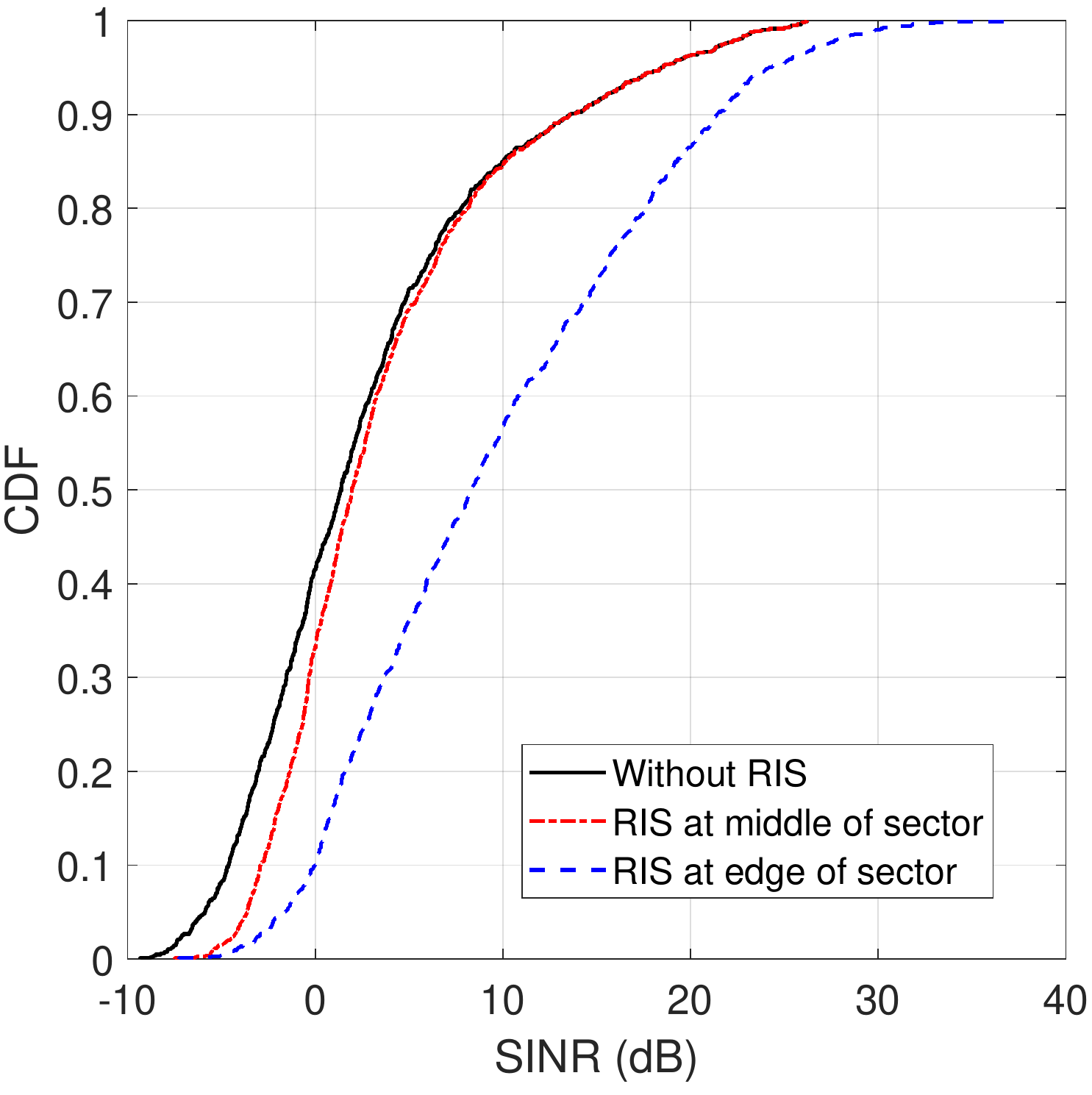}
\caption{CDF of SINR, all UE at cell edges }
\label{fig:fig_simulation_2_SINR}
\end{figure}

In this case, two deployment modes of RIS deployment are simulated, respectively.
In Fig. \ref{fig:fig_layout_UEedge_RISmiddle}, RISs deploy in the middle of the cell.
In Fig. \ref{fig:fig_layout_UEedge_RISedge}, RISs deploy at the edge of the cell.
Fig. \ref{fig:fig_simulation_2_Pt} and Fig. \ref{fig:fig_simulation_2_SINR}
show the CDF of the received signal power $P$ and SINR, respectively.
It can be seen that RIS at edge of sector can achieve better performance rather than the RIS in the middle of sector.

\subsection{Simulation Analysis}
From the simulation results, it can be concluded that
\begin{itemize}
\item Deploying RISs in wireless networks can significantly improve the performance of system.
\item Compared with deploying in the middle of the cell, it is preferable to deploy RIS deployed at the edge of the cell in terms of performance enhancement.
\item Either large number of elements per RIS or large number of RIS per sector can improve the the performance of system.
\end{itemize}


\section{Conclusion}
In this paper, we have investigate the channel model of RIS assisted network and proposed a new antenna model to be incoordination with the characteristic of RIS.
With our proposed model, system-level simulations of network performance under various scenarios and parameter configurations are performed, followed by some insights.

\bibliographystyle{IEEEtran}

\bibliography{Thesis_RIS}

\end{document}